\documentclass[11pt]{article}
\usepackage{graphicx}
\usepackage{amsmath}
\usepackage{geometry}

\input{tcilatex}

\begin{document}

\author{\textbf{Scott M. Hitchcock} \\
National Superconducting Cyclotron Laboratory (NSCL)\\
Michigan State University, East Lansing, MI 48824-1321\\
E-mail: hitchcock@nscl.msu.edu}
\title{\textbf{Time and Information}: The Origins of 'Time' from Information Flow
In Complex Systems}
\date{Presented: Friday, June 29, 2001}
\maketitle

\begin{abstract}
The \textbf{problem of time }can be solved in principle by taking the
viewpoint that information created by unstable physical systems or \textbf{%
Feynman Clocks} (FCs) is transferred by signals to detectors as \textbf{%
infostates} and then used to \emph{compute} time \cite{R1} using a new
invention, the \textbf{T-computer}. This computed 'time' is used to define
the time coordinates for events in space-time maps. The direction and
dimension of 'arrows of time' follow from the ordering of the numbers used
to label event 'times'.
\end{abstract}

\section{Introduction}

\begin{quotation}
''Why is the flow of psychological time identical with the direction of
increasing entropy? The answer is simple: Man is part of nature, and his
memory is a registering instrument subject to the laws of information
theory. The increase of information defines the direction of subjective
time. Yesterday's experiences are registered in our memory, those of
tomorrow are not, and they cannot be registered before tomorrow has become
today. The time of our experience is the time which manifests itself through
a registering instrument. It is not a human prerogative to define a flow of
time; every registering instrument does the same. What we call the time
direction, the direction of becoming, is a relation between a registering
instrument and its environment; and the statistical isotropy of the universe
guarantees that this relation is the same for all such instruments,
including human memory.'' -Hans Reichenbach\cite{R2}
\end{quotation}

What is 'time'? Is 'time' a 'dimension' similar to, but yet clearly
different from, the three 'dimensions' we associate with space? What is the
relationship of 'time' to consciousness? The purpose of 'time' is to allow
us to understand the patterns of change in the locations and configurations
of 'things' in space. We do this to predict where they are going, how they
are going to get there, and in what forms they might take in the 'future'.
We also want to know how things evolved from other configurations and states
with respect to a 'clock'. This is how we define the 'past' and 'time' in
'history'.

The problem with this kind of 'time' is that it is coupled to 'space' in
such a way that our notions of changes 'in time' are entangled with changes
'in space'. In order to know that things change we need information. There
are many kinds of 'change' possible for a system. We can use the directions,
velocities, and accelerations of things moving in space to define 'external'
or relative location 'change' and thus a direction 'in time' parallel to the
motion. We can also use the information generated by 'internal' physical,
chemical or information processing logic to define 'change'. We may have a
hierarchy of changes in systems. We choose which behavior or property to
focus on. We can ignore the other behaviors of the system that don't change
by comparison or that do change but are not dependent on the ones we have
selected for observation. It would be possible to say that some things about
a system change and others don't. If we define time by change, then a system
can paradoxically be changing and not changing 'in time'.

We see that a time paradox can occur when we think of this system as having
some properties that are 'changing' and some that are 'not changing' in
time. The resolution to this and other time related paradoxes lies in seeing
that 'time' is a construction applied to those selected aspects of reality
that create or modify information. We know that they 'change' only by
detecting and processing the signals transporting information generated
during transformations from one configuration to another one. These
transformations may appear to be reversible to us if they are in a state
similar to, and perhaps indistinguishable from, a 'previously observed'
state. Reconstructing a configuration of a system is not the same as time
reversal since work must be done on the system. This information is provided
by other systems in the environment such as observers setting up experiments.

All unstable configurations transform irreversibly into more stable ones
producing signals. I suppose this is how we define 'unstable'. This may be
an internal process for isolated systems. Reversibility of a process
requires 'external' information or energy to act on a state to transform it
into a configuration we associate with one we have observed, time labeled,
and stored in some sort of memory.

Are the irreversible and reversible changes we observe in the universe
around us occurring 'in time' or are they creating 'time'? By treating time
as a 'dimension' of space-time we lose some of the information about the
systems involved. The loss of information occurs when we reduce complex
patterns of 'change' into observables. Time's direction and dimension are
generally assumed to have fundamental roles like space in our descriptions
and models of the evolving universe. Paradoxically, 'time' is not an \emph{%
observable} like energy, position, and momentum. Without 'change' we could
not define time. We would be stuck in a circular set of arguments if \emph{%
change defines time} and reversibly, \emph{time defines change}. If we
assume that 'change' can occur without 'time' then we can examine how 'time'
is created by our observation of 'change'.

This is the essence of the 'problem of time'. One approach to solving this
problem is to find a way that 'time' can be 'constructed' or 'derived' from
a time-independent framework. 'Time' may be a 'construction' or 'map' made
by information processing 'systems' like us and 'mapped' back onto a
'timeless' but changing' universe. We will explore this approach in this
paper.

\section{The 'Problem of Time'}

The 'problem of time' is more that just what is the nature of 'time'. The
real 'problem of time' is that we believe it is a fundamental property of
the universe in the same sense that space is. We know that 'time' is clearly
different from space but yet we force them together into space-time.
Space-time is a very useful construction for understanding physics.

Why does there appear to be reversible and irreversible properties
associated with systems that change or evolve depending on their scale and
degree of complexity? Time is first and foremost a tool for ordering events.
It is a number that is computed from information \emph{read off clocks} and
then used to label 'observed' information representing the events in some
sort of memory. The computation of time involves the creation of \emph{time
differences} represented by these numbers from clocks.

The key to understanding the fundamental nature of 'time' is that signals
form causal relationships between standard clocks and observed events.
Unstable systems are the source of signals and the 'information' that we use
to construct the 'measure' of change that we call 'time'. In order to
construct 'time' as a measure of change for reconfigurations of matter in
the universe, the conversion of the energy, spatial (directional)
information, and matter carried by signals into the 'time numbers' connected
with states of standard clocks, differences between time labels must be
computed. This computation process also defines \emph{arrows of time}
pointing between the time labels.

The generalization of a local model of time to cosmological systems such as
Quantum Growing Networks \cite{R11} and other network theories applied to
the evolution of the universe may be illuminated by understanding the role
of the T-computer at various hierarchical scales. The hierarchy of scales
range from trans-Planckian primordial nodes to cosmological expansion to
investigations into 'clocking' mechanisms related to microtubule and DNA
'T-computers' in cells scaling hierarchically upward to the sense of 'time'
and its role in defining consciousness in terms of the minds awareness of,
or \textbf{attention} to 'change'. In this paper, '\emph{attention}' refers
to the focused or 'directed' computational activities of a causal network or
computer engaged in processing specific \textbf{infostates} in the gates or
nodes of selected causal network circuits or pathways.

\section{Information: The Source of 'Time'}

What do we mean by 'information'? We will use a general definition that
encompasses the definitions used in quantum computation and classical
information theory. Information is any kind of 'label' attached to a
physical state of a system. These labels are abstractions created by
processing various forms of signals originating in or scattered off an
observable system. The source of information is always found in signals.
Signals may take the form of particles, photons, atoms, molecules,
electromagnetic waves, images, sound waves, and electrical currents for
instance. Signals can carry information specific to their sources such as
transition energy changes, direction, motion, and type of source. Some of
the information content of signals is found in the collective
characteristics of various types of signals from the source. An example of
this is the 'spectra' of hydrogen. Each emitted photon tells us about a
specific transition or state, but all the different photons forming the
'spectrum' tell us about the complex quantum structure of hydrogen as a
'system' with many possible configuration states or 'infostates'. There are
many other examples of collective behaviors representing states of systems
that are more that the mere sum of the various separate signals that can be
generated by reconfiguration processes in these sources.

First we look at a general conception of the infostate as packets or
ensembles of information that 'travel' together as a single object, the
infostate. An infostate is an extension of the standard interpretation of
the wave function for a system representing all the known information about
the state of the system. This includes the standard primary or first order
observables such as energy, position, and momentum. It also includes
secondary or second order observables such as 'images' (e.g. collective
surface information), shape, (geometric configurations), material
composition and distribution, internal kinetics and motion, chemical
processes, internal 'logic', or information processing activities and
algorithms, etc. For a system, $S$, the infostate is encoded as a 'qword'
with 'qubit' entries or other higher order qwords (expanded infostates with
additional information acquired by logic operations at various gates or
nodes in the computer) representing the primary and secondary information
together but as an infostate 'in' a causal network. For example an infostate
may be characterized by $I(S_{n})$ for a given configuration '$n$', where: 
\begin{equation}
I(S_{n})=I((n,E,m,p),(\text{first order terms}),...,(\text{second order terms%
}),...,etc.)
\end{equation}

The second order terms result from the observer's addition and expansion via
entanglement (direct product) of complex secondary information to the
primary infostate originating in the source signals. In other words, the
expansion of the qword size by the number of qubits of secondary observables
representing the net or collective active infostate in the T-computer or
other causal network.

The size (e.g. the number of qubit entries in the infostate qword) and
content (e.g. states of the individual qubits in the infostate qword) of
infostates can be modified by logic operations on them as they propagate
through causal networks. The creation of new information and infostates is
consistent with a 'conservation of total information law' for the entire
universe \cite{R5}. By this we mean that overall the emergence of new
infostates as the result of quantum 'logic' (physical) operations on an
active infostate of a node or gate in an information processing network.
Where does this information come from? The creation of new information (the
reverse process of 'information loss') may be the result of 'decoherence'
processes for unstable systems at the interface (event horizon) of matter
with the vacuum. This seems consistent with recent work by \emph{Claus
Kiefer ('Hawking radiation from decoherence' LANL Archive gr-qc/0110070, 15
Oct 2001)} for black holes in which he has found that there is no
information loss paradox. If the entanglement entropy of the universe is
greater than or equal to the conventional entropy, the vacuum is the logical
'reservoir' for the additional 'information' observed in evolving complex
systems. The conversion of the entanglement entropy 'stored' in the non-zero
energy density of the vacuum into 'observable' information requires further
investigation.

Information is represented by the individual qubit entries forming qwords
that describe more than one property of the detected 'signal'. In the
context of quantum computation, qubits are localized infostates carrying
information such as position, momentum, energy, spin states, polarization
states, or any other quantum 'observable'. In order to keep these various
qubits 'together' as a qword, physical quantum 'registers' or memories are
used to allow parallel transport of the collective qword infostate through
causal networks. Causal networks are essentially computers. They may be
quantum or classical computers depending on the physical nature of the
information and the 'logic gates' involved in processing and storing the
infostates. In the information model of time \cite{R1}, a process of
observed and standard clock signals are paired so that the observed signal
can become a 'time labeled' infostate in an ordered set of 'infostates'
stored in the 'memories' or registers of the observer.

'Classical' information such as the binary states in semiconductor
computers, collective representations of thermodynamic coordinates like
temperature and pressures, and Shannon's definition based on entropy, are
examples of hierarchical mesoscopic and macroscopic states built from
microscopic quantum infostates of the atomic and molecular components acting
collectively as a single system. The concept of collective excitations
represents a bridge between quantum and classical descriptions of complex
systems. Collective excitations or infostates indicate plateaus of
complexity (POCs) in complex systems in which new behaviors can emerge. The
computation of the 'times' associated with 'events' allows us to construct
time ordered sets of events and 'arrows of time'. These secondary
information structures or maps can be used to define the causal
relationships between infostates (e.g. 'memories').

\section{'Computing' Time from Information}

If time is a number that is computed from information, how is it created and
how is it realized as information that can be used by the observer to give a
time coordinate to an event? Time can be thought of as a form of information
about causal relationships of events constructed from signals originating in
an observed system. These constructions are sequentially ordered 'maps' that
are built from sets of events as the result of the computation and
application of 'time labels' to the information states produced in the
observer or his/her 'equipment'. The 'computation' process for generating
these labels uses signals from a standard clock paired with the signals from
observed events. This allows us to define the cause and effect relationships
between the events in our environments.

In general, the observed system (the source of 'signals') is a kind of
'clock' called a Feynman Clock or FC \cite{R1}, \cite{R3}, \cite{R4}. A
Feynman Clock is any unstable quantum system that decays into another state
or configuration of the system or into sets of 'decay products' such as
those created in high-energy particle collisions.

The motivation for this 'time' theory follows from the use of Feynman
Diagrams in particle physics. 'Time reversibility' or time symmetry is
usually taken for granted at the microscopic scale where Feynman Diagrams
are useful. The process reversal of particle interactions is not the same as
time reversal since it is the transient excited state of the composite
system of incoming particles that decays irreversibly whether or not the
incoming particles are exchanged for the outgoing ones and vice versa. The
'recreation' of an unstable state of a system is not the same as going 'back
in time'. Since time is created by reconfigurations of unstable systems, the
direction of 'time' can only be defined by the direction of 'information
flow' or 'infostate' propagation from sources, via signals, to the detectors
and 'logic' gates (or nodes) forming 'causal networks'. At the quantum
scale, causal networks can be thought of as generalized quantum computers
capable of processing a wide spectrum of signals and their energies (e.g.
electronic interactions with photons in many-electron atoms) expanding the
computing capabilities of matter beyond binary state (e.g. 0's and 1's).

Any system in an unstable state, processes incoming information into
outgoing information upon it's decay, decoherence or reconfiguration. The
apparent 'time reversibility' of fundamental interactions is due to
intervention upon the system putting in into a state that the observer
considers the same as some 'past' state by comparison of the configuration
of this 'process restoration of some reference state. This requires
information in the form of a signal that can be detected by the system and
converted into a 'new' configuration similar to a 'past' one with respect to
the observers' clock.

The theory of time as 'information' is general and compatible with all
current established physical theories. It can be applied to any physical
system or theoretical model including second order constructions of time
whose 'dimensions' or 'directions' use complex numbers or multidimensional
time 'coordinates'.

Any complex network of physical objects that involves signal generation,
signal detection and subsequent 'processing' of induced infostates in the
gates, nodes or devices forming the network, can be understood within the
context of this theory. This includes examples such as the particle
accelerator systems involved in observations of subatomic particle
collisions, complex information processing systems such as optical image
formation by eyes and subsequent image processing by brains in living beings.

The infostate contains all relevant 'observable' information for the system.
You can think of this infostate as a sort of 'word', in the computational
sense, in which each of the '$n$' entries or bits (qubits etc.) form an '$n$%
'-tuple. One may use some or all of the information bits or qubits etc. in
an infostate for a given purpose or operation. These entries would include
information in the conventional sense such as energy, momentum, wavelength,
etc. and in a broader sense; 'images', material composition, internal logic
operations or physical transitions, etc. There may be a conservation of
information 'law' concerning the abstract 'magnitudes' of infostates with
respect to the entire universe \cite{R5}, but the infostate paradigm
suggested in this paper is applicable regardless of global information
creation or loss (e.g. black holes as logic gates).

\section{Feynman Clocks, Signals, and Detectors in Causal Networks}

Unstable systems or Feynman Clocks (FCs) create Signals carrying
'Information' away from the source to other clock or detector systems in the
process of reconfiguration (or decay) to more stable states. Systems of
permanent or transient sets of FCs and the signals between them, form causal
networks which are the basis for the computation and subsequent creation of
the direction and dimension of 'time'.

The 'time' differences associated with the reconfiguration of a system are
computed using initial and final state signals. A complex system may have
many overlapping reconfiguration processes at work with different
'lifetimes' for each one. 'Time' can be thought of as the state information
representing the configuration 'differences' between states of the observed
system as labeled by the signals from a standard clock. The process of
pairing observed signals with standard clock signals allows the computation
of 'elapsed time', 'lifetimes', and relativistic time contraction and
dilation effects. The ordering of these 'time labeled' events with respect
to the set of real numbers provides the basis for the direction and
dimension of 'arrows of time' at all levels of complexity.

\section{From Feynman Diagrams to Feynman Clocks}

We begin with the example of the conversion of a Feynman Diagram \cite{R6}, 
\cite{R7} of a fundamental interaction into a generalized model I call the
Feynman Clock. For example, the strong force or interaction involved in the
scattering of an incoming proton and neutron is mediated by a $\pi$-meson
with an approximate maximum nucleon separation distance $d_{FC}=1.5X10^{-15}m
$ or the approximate 'range' of the strong force (see \textbf{Figure 1}).

Using the Uncertainty Principle, we see that a $\pi$-meson can come into
existence by violating energy conservation by an amount of energy given by
the following relation:

\begin{equation}
\Delta E=(m_{\pi-meson})\times c^{2}
\end{equation}

The box in the space-time diagram below represents the transient Feynman
Clock (FC). The space-time 'size' of the Feynman Clock is ($%
d_{FC}\times\Delta t_{FC}$) or about $7.5\times10^{-39}m\times sec$. Where $%
\Delta t_{FC}=5\times10^{-24}s$ is the 'lifetime' of the Feynman Clock. This
is also $\pi $\textbf{''The Direction of Time''} by \emph{Hans Reichenbach},
Edited by Maria Reichenbach, Dover Publications, Inc., Mineola, NY,
1999.-meson signal 'transit time' mediating the strong interaction and
causing the reconfiguration or decay is accompanied by the production of two
signals in the form of new proton and neutron trajectories. Now we make an
intermediate step towards the causal network representation of a Feynman
Clock 'node' or 'gate' component by observing that the 'information flow'
through the target space can be viewed as a time-independent map in an
'info-space' diagram illustrated in \textbf{Figure 2}.

The 'info-flow' diagram for the proton-neutron creation of a Feynman Clock
above can now be represented by a causal network' diagram in which the nodes
are Feynman Clocks. Sets of these Nodes and the sets of signals connecting
them form the Causal Networks ('wiring') that maintain 'order' in complex
systems. (Authors note: The ground state or 'signal detection' states of a
Feynman Clock may be referred to as the 'detector' mode of a FC or as a
'Feynman Detector'). All systems capable of detecting incoming signals,
processing them, and producing outgoing signals represent the general form
of a Feynman Clock. The causal network node representation of the above
space-time and 'info-space' diagrams is illustrated in \textbf{Figure 3}.

Why invent the 'T' Computer model? The T-computer creates 'time' labels and
causal and temporal relationship between events and computes the 'time' that
we use in everyday life. It also allows us a chance to see how 'time' can be
constructed from a 'timeless' space in which things evolve. Time is a
'construction' derived from the applications of logic, ordered sets, and
standard clocks that allow us to locate events in spatial and temporal maps.
These maps are the source for the 'dimension' and 'direction' for 'arrows of
time'. Transient, permanent, and adaptive wiring of network circuitry can
occur in sets of logic gates, signals, shift-register-clocks, and memories
and clocks can drive information flow or chart its progress.

\section{The T-computer: a 'Solution' to the 'Problem of Time'}

In order to 'solve' the problem of time we must see how 'time' is computed
in this model. It would be interesting to see how the direction and
dimension of time are also computed. The basic computation of a 'lifetime'
or 'elapsed time' involves a physical information processing system I call
the T-computer. This could be the observation of the creation of a Higgs
particle in a very complex and very large accelerator acting as a
macroscopic quantum T-computer or the detection of an ancient photon from a
distant galaxy by the rods or cones in the retina of your eye. All 'T'
computations involve pairing signals from two or more events with coincident
(with respect to the observer) standard clock signals (from atomic clocks to
the heartbeat). They also require some sort of 'logic' that can compute
differences in the 'time labels' assigned to the information states
representing the 'observed events'. This is how 'time' is created as a
'secondary map' (e.g. events in space-time representations) of change in an
evolving universe. This concept can be seen in complex biological
computations of time perhaps in microtubule quantum computer components of
the neurons forming the increasing complexity of the hierarchically scaled
'classical' computers of mesoscopic and macroscopic neural networks in the
brain. The computation of 'time' via quantum and classical T-computers is
essential for the existence of consciousness as a measure of our interaction
with our environment. Some of the tools of quantum computation \cite{R8}, 
\cite{R9}, \cite{R10} have been adapted to illustrate this approach.

\section{T-Computer Principles}

The progression of information transfer through the network is mapped by the
position of the infostate representing the coupling of the original
coincident signals and subsequent 'processed' or computed infostates as they
are operated on by sequential 'gates' in the network. This information
propagates through the T-computer from the Feynman and Standard clock
sources to a time labeled memory. The final calculation of the 'time
difference', $\Delta t=t_{2}-t_{1}$, between any two observed events or
infostates stored in two different memory locations requires physical
'logic' that can find the 'difference' between the time labels associated
with the stored event information. In the following equations the nth
composite state, $S_{n}$, is listed for the entire T-computer acting as a
single quantum system (\textbf{see Figures 4 and }5). This state represents
a given configuration for the entire system focusing on the 'active'
infostate in the causal network. In the following, $\left|
Network\right\rangle $, refers to the 'inactive' collective state of the
network components not involved with the location of the 'active' infostate.

The sequence of information flow through a causal network (e.g. T-computer)
is tracked by 'active' or 'excited' (e.g. $\left| FC\ast\right\rangle $
infostate representing the spatial location in the network of all the
relevant information corresponding to the observed event. The non-active
states of the remaining components in the network are summed in the 
\TEXTsymbol{\vert}Network0\~{n} term. The nth (computational) state of the
entire T-computer system is $S_{n}$. This corresponds to the infostate at a
physical gate location in the causal network.

We will assume that the initial configuration of the composite system is
given by $\left| S_{n}\right\rangle =\left| S_{0}\right\rangle $, for $n=0$.
The standard clock and the 'observed' system provide the information
(signals) to the T-computer detectors through open space or by closed
'circuits' or 'guides'.

We want to remember that we implicitly use another distinct T-computer when
we assign a 'start time' for the flow of information through any causal
network or T-computer. Since the point here is that 'time' is a bit or qubit
of information generated by the T-computer, the start and stop times can be
ignored. We will also ignore any 'decoherence' effects on the infostate
through its' interaction with the environment (e.g. vacuum) since the decay
'lifetime' of a system due to this coupling is already included in the
system start and stop signals processed by the observer. The network will be
considered to be robust enough to withstand decay of the local infostate in
the active gate or node in the network.

\section{The Flow and Contents of 'Infostates' in the T-computer}

The equations below are intended to illustrate the general features of the
T-computer. The details of the physical 'logic' gates and nodes forming a
T-computer and the appropriate equations describing them are currently being
investigated by the author and will be published at a later date.

We assume that we have an initial configuration of the entire quantum system
involving the signal sources, however extended or remote in space, and the
causal network forming the T-computer given by the superposition (sum) of
all the infostates of the components in the extended system. These are
represented by the collective state of the system, $S_{0}$, defining a
reference 'start' state of the T-computer where the state index is $n=0$.
The following equations represent the state of the entire system in which
the 'active' signals, nodes or gates are identified separately from the
remainder of the 'inactive' components of the network. This allows us to see
how the 'infostate' is created and propagates through the network. We can
see how it changes ('expands' with additional qubits into larger qwords or
is 'reduced' to the selected information needed for a time computation) as
it is acted upon by the 'logic' of the network.

The active components (e.g. signals, 'nodes' or 'gates') of the T-computer
system (i.e. the 'current' location of relevant information from the
'observed and time labeled' expanded infostate including additional qubits
form any extra information created by processing or logic activities on the
incoming infostate) and the remainder 'inactive' network for the start state
is given by states in Dirac notation.

We begin with the initial state of the T-computer:

\begin{equation}
\mathbf{S}_{0}:\left| FC^{\ast}\right\rangle +\left| D\right\rangle +\left|
SC^{\ast}\right\rangle +\left| Network_{0}\right\rangle
\end{equation}

The active (excited) states of the FC and SC are direct product states of
their respective 'ground' states and the 'potential' outgoing signal. They
are initially entangled until full decay at which point they become a
'distinct' linear superposition of quantum systems. When the source and its
signal become 'measurably distinct' we can consider them to be 'classically'
separated. This 'separation' is represented as a sum ($+$) rather than the
direct product ($\otimes$) .

\emph{The causal network forming the T-computer is a 'quantum' system.
'Classical properties' of a network are the result of mesoscopic or
macroscopic collective excitations or behaviors resulting from the composite
collective interactions of the quantum sub-systems or sub-networks. The
distinction between quantum and classical scale phenomena is mainly a matter
of the observers' choice of what is to be observed. For example, one could
measure the 'classical' temperature of a gas cell (a 'thermodynamic'
infostate) or the optical scattering of individual 'quantum' photons by the
particles in that same gas cell (quantum causal network infostates). The
total system is neither quantum or classical alone, but a system with both
quantum and classical infostates at various hierarchical levels of
complexity.}

\emph{\ We note that the Dirac notation lends itself to an oversimplified
'compact' representation of the state of a system considered to have a
distinct quantum identity. We neglect the interactions of components of the
system with each other or the environment if they are not 'active'. 'Active'
components are defined by the 'attention' interaction of the observer with
the T-computer. This interaction is the result of an ongoing coupling or
feedback between the information source, a standard clock and the observer. }

\emph{\ The superposition of the quantum states or the quantum components of
an ensemble system represents a system that does not support or generate
collective excitations as the result of a 'condensate' state. The condensate
state of an }$\emph{n}$\emph{-body system that generates or supports
collective excitations is the result of the coupling of all the components
in such a way that they are more than a superposition. The expression of
this collective or entangled condensate is the direct product of the system
states whose whole is clearly more that the sum (superposition) of the parts.%
}

\begin{equation}
\mathbf{S}_{1}:\left[ \left| FC^{\ast}\right\rangle \longrightarrow\left|
FC_{0}\right\rangle \otimes\left| \lambda_{FC}\right\rangle \right] +\left|
D\right\rangle +\left[ \left| SC^{\ast}\right\rangle \longrightarrow\left|
SC_{0}\right\rangle \otimes\left| \lambda_{SC}\right\rangle \right] +\left|
Network_{1}\right\rangle
\end{equation}

The decay rates of the FC and the SC may be different. We will assume that
the detector will 'hold' the FC infostate until the next available signal
arrives from the standard clock. When both signals are detected they are
converted into a 2-qubit infostate.

\begin{gather}
\mathbf{S}_{2}:\left[ \left| FC_{0}\right\rangle \otimes\left| \lambda
_{FC}\right\rangle \longrightarrow\left| FC_{0}\right\rangle +\left|
\lambda_{FC}\right\rangle \right] +\left| D\right\rangle \\
+\left[ \left| SC_{0}\right\rangle \otimes\left| \lambda_{SC}\right\rangle
\longrightarrow\left| SC_{0}\right\rangle +\left| \lambda_{SC}\right\rangle %
\right] +\left| Network_{2}\right\rangle
\end{gather}

The collective state of the total system is one in which the 'signals' are
in transit to the detector while the rest of the network is in its 'ground'
or signal detection 'ready' configuration. Once the 'classical' or spatially
distinct signals become close enough to their targets to become superimposed
and then entangled with the detectors, their identity as quantum signals
traveling 'classically' through space is destroyed. They are now coupled to
the detectors as indicated by the direct product of the infostates of the
active front-end components of the T-computer.

The 'inactive' components of the network including the FC source and the SC
'time pulse generator' are lumped into the superimposed subset of the
systems causal network as represent in the last term below. The 2-channel
detector is now in a collective excitation state in which two qwords (or
qubits) are stored in parallel quantum registers.

\begin{gather}
\mathbf{S}_{3}:\left[ \left| D\right\rangle +\left|
\lambda_{FC}\right\rangle +\left| \lambda_{SC}\right\rangle
\longrightarrow\left| D\right\rangle \otimes\left| \lambda_{FC}\right\rangle
\otimes\left| \lambda_{SC}\right\rangle \right] \\
+\left[ \left| FC_{0}\right\rangle +\left| SC_{0}\right\rangle +\left|
Network_{2}\right\rangle \longrightarrow\left| Network_{3}\right\rangle %
\right] \\
=\left| D\right\rangle \otimes\left| \lambda_{FC}\right\rangle \otimes\left|
\lambda_{SC}\right\rangle +\left| Network_{3}\right\rangle
\end{gather}

The composite infostate of the detector is an expanded 'qword' with two
qwords concatenated into a larger qsentence. The qsentence infostate can now
be propagated along the network for used as a 'time stamp' for the 'event'
originating in the FC and labeled by the SC.

\begin{equation}
\mathbf{S}_{4}:\left[ \left| D\right\rangle \otimes\left| \lambda
_{FC}\right\rangle \otimes\left| \lambda_{SC}\right\rangle =\left| D^{\ast
}\right\rangle \right] +\left| Network_{3}\right\rangle =\left| D^{\ast
}\right\rangle +\left| Network_{4}\right\rangle
\end{equation}

The detector infostate is now given a 'label' that will be used later to
compute the 'elapsed time' between events in memory or the coordinate 'time'
used in standard space-time.

\begin{gather}
\mathbf{S}_{5}:\left[ \left| D^{\ast}\right\rangle \otimes\left| \tau
_{n}\right\rangle \longrightarrow\left| M_{k},t_{k}\right\rangle +\left|
D\right\rangle \right] +\left| Network_{4}\right\rangle \\
=\left| M_{k},t_{k}\right\rangle +\left[ \left| D\right\rangle +\left|
Network_{4}\right\rangle \longrightarrow\left| Network_{5}\right\rangle %
\right] \\
=\left| M_{k},t_{k}\right\rangle +\left| Network_{5}\right\rangle
\end{gather}

At this point a signal from another memory location carrying the 'time
label' information is shifted into the comparator.

\begin{equation}
\mathbf{S}_{6}:\left[ \left| M_{k+m},t_{k+m}\right\rangle +\left|
M_{k},t_{k}\right\rangle \right] +\left| Network_{6}\right\rangle
\end{equation}

The 'time labels' are qubits in the $n$-bit word representing the infostates
for the processed information from the source and standard clock 'events'
stored in the various addressable memory locations.

\begin{equation}
\mathbf{S}_{7}:\left[ \left| M_{k+m},t_{k+m}\right\rangle \otimes\left|
M_{k},t_{k}\right\rangle \right] +\left| Network_{6}\right\rangle
\longrightarrow\left| M_{k+m},t_{k+m},M_{k},t_{k}\right\rangle +\left|
Network_{7}\right\rangle
\end{equation}

The interaction of the infostates from the two memory locations takes place
in a logic gate that compares the 'time' qubits via a 'subtraction'
operation. This follows from the same kind of logic used in conventional
computers that address specific bits needed for a logic operation involved
with finding time differences for time labeled events.

\begin{gather}
\mathbf{S}_{8}:\left| M_{k+m},t_{k+m},M_{k},t_{k}\right\rangle +\left|
Network_{7}\right\rangle \\
\longrightarrow\left| M_{k+m},M_{k},\left( t_{k+m},t_{k}\right)
\right\rangle +\left| Network_{7}\right\rangle \\
=\left[ \left| M_{k+m},M_{k}\right\rangle \otimes\left|
t_{k+m},t_{k}\right\rangle \longrightarrow\left| M_{k+m},M_{k}\right\rangle
\otimes\left| \Delta t_{n}\right\rangle \right] +\left|
Network_{8}\right\rangle
\end{gather}

The processing of the time labels for infostates corresponding to two events
results in a 'time' infostate whose information 'content' is the difference
between the labels. We are assuming that real numbers are used here, but
complex numbers or any other set of number-like labels may work as long as
the physical states of the register in which the computed 'time' can be
translated into other higher order languages that 'interpret' causal
relationships between events.

\begin{gather}
\mathbf{S}_{9}:\left[ \left| M_{k+m},M_{k}\right\rangle \otimes\left| \Delta
t_{n}\right\rangle \right] +\left| Network_{8}\right\rangle \\
\longrightarrow\left| \Delta t_{n}\right\rangle +\left[ \left|
M_{k+m},M_{k}\right\rangle +\left| Network_{8}\right\rangle \right]
\longrightarrow\left| \Delta t_{n}\right\rangle +\left|
Network_{9}\right\rangle
\end{gather}

The information encoded in the difference between two time labels for two
events is extracted by logic that can evaluate the absolute value of the
difference between the two event 'times' resulting typically in a real
number. These time label numbers may be physical states such as the number
and polarity of charges, analog voltages, discrete binary sets of voltages,
polarization states, or spin states. The numerical difference between two
time labels corresponds to the physical difference between their physical
states in the tubit location of the infostate qword. The physical comparison
of two states by the logic of the gate in the T-computer results in the
following creation of another physical state in a register that can be
translated into a number by higher order information processing:

\begin{equation}
\mathbf{S}_{10}:\left| \Delta t_{n}\right\rangle +\left|
Network_{9}\right\rangle \longrightarrow\left[ T[\left| \Delta
t_{n}\right\rangle ]=\Delta t_{n}=t_{classical}\right] +\left|
Network_{10}\right\rangle
\end{equation}

Where 'I' is the computers 'time infostate' resulting from the computation
of the time label differences in two event infostates by the time operator,
T, acting on the two-qubit infostate, $\left| \Delta t_{n}\right\rangle ,$
extracted from two memory locations. The action on this state results in
conventional time, $T[\left| \Delta t_{n}\right\rangle ]=\Delta t_{n}$. This
is the time difference 'magnitude' between events stored in memories $(k)$
and $(k+m)$ with respect to a standard clock.

The \textbf{constructed} \emph{equation of time representing the}\textbf{\
bridge }\emph{between quantum and classical processes is:}

\begin{equation}
T\left( \left| \Delta t_{n}\right\rangle \right) =\Delta t_{n}=t_{classical}
\end{equation}

'Time' differences and the information defining the order of infostates
representing the observed events can be used to create temporal pointers or
'arrows of time' between 'earlier' and 'later' infostates (i.e. $(k),(k+m)$%
). This is the 'output' of the T-computer.

The magnitude of the differences in the time labels along with the 'pointer'
are used to construct arrows of time and the 'dimension' and 'direction' of
the time axis in standard $(3+1)$ space-time. This is the 'classical' time
that is generally used as the time 'variable' in standard physics equations
of motion.

The real number time differences coupled with the loading of info-states in
memory locations, $M_{k+m}$ and $M_{k}$ along with the set of all ordered
events defines for the observer a 'timeline', time 'direction' and time
'dimension' (usually = '1') coupled to a standard 3-space resulting in a $%
(3+1)$ space-time. It also defines a \textbf{Quantum Arrow of Time (QAT)}
for signal creation and induced infostates in detectors originating in the
irreversible reconfiguration of unstable 'excited' states. \textbf{%
'Classical Arrows of Time' }or\textbf{\ CATs} are built on collective or
generalized information flow in composite quantum systems acting with
behaviors that can be described by classical equations of physics and
pointing from unstable system configurations to more stable ones.

Quantum information encoded as qubits and qwords are the 'contents of
'infostates' resident in gates, registers and memories. Collections of
memories can support many qwords as single information objects. They may be
extended quantum objects with serial (\emph{sequential excitation network or
SEN}) properties or collective properties (\emph{collective excitation
networks or CENs}). Combinations of qwords acting like a single quantum
object can form qsentences. These various information structures are
physical 'infostates' of the signals or gates in which they reside. The
transfer of physical objects combined with their information content defines
causal networks and T-computers. The information flow in the universe occurs
without any explicit dependence on time as a 'dimension'. Real and complex
numbers, or any ordered set of objects can be used to 'time label' events.
Unusual units of time can also be created to represent causal relationships
in theoretical physics models where complex processes involve 'mixing' of
spatio-temporal information as long as their 'ordering' is understood.

Recognition that 'time' is created by complex systems capable of 'computing'
it, may clear up 'time' related paradoxes and issues related to causality,
information theory, and the 'experience' of time inside complex states of
'consciousness'.

\section{ Do T-Computers Already Exist?}

The primary motivation for creating a T-computer model is to provide a
conceptual basis for understanding 'time' is a computational artifact
resulting from physical 'changes' in the configurations of matter in the
'time-independent' space of the universe. Is this a realistic approach? We
have seen that it is very difficult to find a primitive concept of time that
doesn't implicitly assume 'time' as a fundamental dimension of space. If we
suspend our belief that all change occurs 'in time', then we need to
understand how 'time' arises as a measure of change without assuming what we
are trying to show (the classic petitio principii logic fallacy). We know
that if nothing changes then 'time' has no meaning. If things change then
there is information transferred to other things. If this information can be
used to understand patterns of change, then ordering of the information into
cause and effect relationships between events requires a sense of 'time'.
This sense of time is an effect of computing time labels with respect to
some clock either external (e.g. 'atomic clocks') or internal (e.g. 'heart
beats') to the observer.

Are there physical systems that actually 'compute' time in the way we have
outlined above? We use 'clocks' everyday and we don't seem to be aware that
we are 'computing' time. We read numbers off these clocks and 'pair' them to
events. The T-computer represents the process by which we assign these times
to events at a fundamental level. Whether the brain operates in a way
similar to the T-computer is an open question. Are there examples of the
T-computer method in other systems?

The answer is yes. All 'detectors' and detector control systems are
information processing systems that 'clock' or coordinate and 'calibrate'
events relative to each other and to the standard clock. All detectors
supplying information to causal networks in the form of simple or complex
infostates are forms of 'T-computers' when the infostates are 'time labeled'
by using standard or internal reference or calibration signals created by
'clocks'. There are many examples from the physical sciences such as
determination of 'lifetimes' of the products from particle collisions in
high energy accelerators, astronomical observations of intensity and
spectral variations in stars, and determination of the expansion rate of the
universe to name only a few. Time labeling in biological systems is
essential for survival. Responding to the motion and activities of predators
and prey of all size scales requires an ability to predict future movements
based on those just observed. At a primitive level this means time ordering
sensory data in order to respond to changes in the animals environment. We
see the interface of instrument and biological T-computers in the medical
monitoring of heart and brain activity. A simple but nearly universal
example of a biological T-computer system is the eye, the optic nerves, and
the time labeling neurological activities in the visual information
processing regions of the brain.

Another possible application of the T-computer concept is in cosmology and
the structure of the early universe. An approach being explored by Paola A.
Zizzi, looks at the universe as a Quantum Growing Network \cite{R11}. The
application of network concepts to evolutionary processes from primordial
configurations of the early universe to activities in the brain will include
hierarchical versions of T-computers wherever change occurs and is
'observed' relative to other systems such as 'clocks'.

\section{T-Computers and Consciousness}

In biological systems T-computers are resident in various hierarchical
components raging from individual cells to organs and the neural networks
that form the brain. The exact physical structure is beyond the scope of
this paper. We can point out one possible example of a quantum scale
T-computer that may be a key element in the large-scale 'collective
excitations' of neural networks that we call 'consciousness'. These are the
microtubules \cite{R12}. The primary function of microtubules in neurons may
be as T-computers that coordinate the processing of sensory information by
the brain. This would give us the 'sense of time' necessary for
consciousness. In the laboratory, T-computers exist as an integral part of
the instrumentation apparatus we invent to extend our sensory range and
therefore our conscious perceptions of the world.

T-computers take on a subjective nature since they may be specific to a
given system whose 'rate of change' is different from other systems. This
subjective or individual nature of time labeling events leads us to
examination of the conscious experience of speeding up or slowing down of
'time'. The T-computer 'clock rate' in biological systems can vary because
of the action of various neurotransmitters that slow down or speed up the
information processing 'speeds' of our neural networks. The subjective
experiences of time dilation or time contraction may be due to variations in
sensory signal sampling rates as well. These rates are a measure of
'attention'. Attention in this context refers to a variable control of the
incoming signal sampling and processing rates.

The rate at which we sample our sense data is compared to our subjective
sense of time. This 'sense' of the differences between each distinct thought
in a sequence or flow of consciousness feels 'constant'. We are unaware of
the non-thought 'time' width between thoughts. The decay lifetimes of each
successive thought may become longer but also the 'downtime' in between can
be longer with respect to an external standard clock.

This prejudice that the external world is speeding up or slowing down, lies
in our belief that our internal reference clock is operating in the same way
that an atomic clock does. Our experience of these time effects by our
consciousness may be due changes in the information processing rates of our
neurons. The rate changes are probably not internal to the neuron but result
from the chemical messenger molecules or the physical action of photons and
macroscopic electromagnetic potentials on neurotransmitter production and
subsequent recovery cycles in synapses. The modification of a T-computers
labeling rates and higher order neural states is similar to changing the
'clock rate' on digital computers. Consciousness may mislead our sense of
the rate of 'flow of time' since our belief is that our internal clock
generates its time labels in a regular and repeatable intervals with respect
to external time reference systems like atomic clocks.

In \textbf{Figure 6} we see how a subjective sense of 'time' or temporal
'attention' to external events is related to information flow from the
observers' environment in the form of light pulse signals from a fixed
frequency standard (e.g. atomic) clock driven source. The square wave
pattern illustrates a sort of on ('1', 'tick') and off ('0', 'tock') state
of consciousness in a stream of collective excitation states. The 'width' of
these states corresponds to the observer's 'attention' or information
processing cycle. While the brain may act asynchronously, when 'attention'
is given to incoming data such as the light pulses from the standard clock,
the T-computer component of the nervous system is engaged and time labeling
of events occurs.

We assume that for this case that one pulse (photon) is emitted for every
'tick' or cycle of the standard clock. A person experiencing (i.e. detecting
and 'time labeling') more events than 'normal' such as the 8 clock signals
in the 'Fast Subjective' frame per internal 'tick', believes that the
external world is moving 'faster' than their memory of a 'normal' rate. For
purposes of this example, we define the normal subjective state to be 4
signals detected per internal 'tick-tock' cycle (each complete cycle has a
'width' corresponding to the 'subjective arrows of time' just below each
figure). Internal clocks calibrate the 'width of each cycle. For organisms,
this might be the heart rate specific to a given species. For devices and
instruments, it might be a mechanical, electrical, or electronic clock.

The person detecting fewer signals (2 clock signals detected per internal
'tick') is convinced that the external world has slowed down. This is a
common experience for people in emergency situations where adrenaline speeds
up their metabolism. In all cases, the observers' reference frame seems
normal while it is the external world appears to change at different rates.
This highlights the fundamental role that an observer or observing
instrument has in defining 'time'. In this example, the T-computer is the
composite system created by the interaction of external environmental
information with the 'internal' clock of the instrument, device, or organism
that processes information.

We note that $t_{F}$ and $t_{N}$and $t_{S}$ all 'feel' like the same
interval to the observer, it is the standard clock that appears to change.
This is the clue to understanding the connection between consciousness and
the creation of 'time' maps like space-time. The term 'subjective' can be
misleading. Systems that time label 'observed' events do so with respect to
some internal or 'subjective' clock. The intervals between events such as
'seconds' are defined by the systems that use them with respect to some
external clock of their choice. The same effect can be obtained if the
observers clock is normal but the signal detection rate is changed so that
for small or imperceptible environmental changes one may sense a slowing of
time. This may be one of the explanations for a sense that 'time drags' when
one is 'bored'.

The relevance of this example is that metabolic rates in biological systems
may determine the T-computer intervals for the time labels. The assignment
of time labels to events by a T-computer can be a local phenomenon connected
to larger and more global clock processes in nature. We can see that this
presents a nice starting point for understanding how the 'psychological
arrow of time' is a construction based on the T-computer properties of
complex neural networks and the collective excitation states associated with
'consciousness'.

\section{'Time Symmetry', 'Time Reversal' and 'Time Travel'}

The apparent time symmetry associated with reversal of processes in particle
physics seems to conflict with irreversible processes in complex systems
made of these particles. This transition from reversible particle
interactions to irreversible ensemble behaviors is due to a misunderstanding
about the relationship of 'time' to information flow. Information originates
in the reconfigurations of unstable systems and 'flows' via signals to other
systems. The key point is that unstable systems represent a source for the
directionality of information flow. This means that if one reverses a
particle collision process then information still flows 'away' from the
unstable system created at the site of the interaction of the particles. Any
arrow of time associated with information flow always points away from
reconfigurations of unstable systems. From this point of view there is no
time symmetry for particle collisions, only process symmetry.

Time 'reversal' is a statement about information flow reversal, not a change
in direction of a fundamental 'dimension' of the universe. The dimension and
direction of time as we use it in everyday life is a construction based on
the interaction of the observed world with our 'minds' and our 'clocks'.
Process reversal is not the same as 'time reversal.

Popular ideas about 'time reversal' would require reconstruction of
infostates of the universe as a whole or at least a sufficiently large local
infostate for 'travel' back in time to an 'earlier' state. Since all local
systems are entangled with the infostate of the universe as a whole we see
that we can only create the illusion of time reversal by construction of a
'set' of configurations of matter that mimic an earlier configuration of an
the irreversibly evolving universe as a whole. At the quantum scale the flow
of information is away from unstable systems. To create an unstable system
incoming information from the environment is required. In this sense time
reversal does not exist.

Time travel has three popular modes of expression. The first is 'backward'
'in time' as discussed above. The second is 'instantaneous' or 'zero elapsed
time' travel across space. The third is travel into the 'future' also 'in
time'. All three require access to information about the 'destination' space
that is assumed to exist 'simultaneously' with the traveler and the problem
is how to transport an observer into one of these non-local infostates. This
requires that past and future infostates of the universe 'exist'
concurrently. The problem here is that previous or past infostates are
'lost' as they are 'computed' into future ones by the dynamics of the
evolving universe at all hierarchical scales of complexity. The 'lost'
(really we mean 'processed') information specific to any 'historical'
infostate means that 'backward' time travel is not possible and future
infostates have not been computed yet.

We compute our future actions in response to past information in our
memories. This allows us to compute 'time' in order to predict 'future'
evolutionary patterns. In this sense we are examples of 'time machines'. Our
ability to 'travel' into the past is just our ability to access memories.
Our ability to 'travel' into the future is our ability to 'imagine'
evolutionary scenarios based on extrapolation of patterns formed by
processing information. It appears that we are stuck in the here and 'now'.

What about 'instantaneous' or 'zero elapsed time' travel across space via
wormholes or some exotic quantum entanglement effect? This may be possible
for 'information' states in the quantum realm, but it appears that the
ordinary real matter we are made of is highly resistant to instantaneous
parallel displacement or teleportation in space.

'Popular' notions about time travel may be misguided at best since they are
the result of our misinterpretations of our constructed maps of time rather
than originating from a deeper understanding of how we create time from
information. This is the source of many philosophical and religious
paradoxes relating to the origin and evolution of the universe where our
'time' map constructions are projected onto the universe as a space-like
'dimension'. We may see that the re-conceptualization of 'time' as a
'construction' or information structure will open doors to a deeper
understanding of the 'changes' in the universe that we associate with 'time'.

\section{The Quantum Computing Capacity of the Universe, Non-locality, and
Achronological Change}

While preparing this paper two very interesting ideas have emerged that
reflect the need for the approach outlined above. First is a computational
limitations for the universe as it 'processes' its evolution \cite{R13}. The
second is that non-local phenomena can occur without a 'chronology' (i.e. a
'before' or 'after') (see \cite{R14}). The computational number of
operations calculated in \cite{R13} implicitly assumes the existence of some
sort of T-computer in order to assign times to computational events or
intervals. This T-computer is integral to the computational activities of
the universe as it calculates its future. The non-local experiments
described in (\cite{R14}) concur with the idea of this paper that 'time' is
a construction and does not exist a priori. The confusion about 'before' and
'after' information transfer in Bell correlations is probably the result of
assuming that quantum non-locality occurs 'in time'. The Bell correlations
may be 'achronological' or 'occurring' without explicit dependence on
'time'. This is consistent with the idea that 'time' is constructed into
temporal maps of causal relationships between events. The numerical values
of to 'time' result from the relative scaling relationship of observed
events calibrated by the observers' chosen standard clock. The constructed
map is then projected back onto 'reality' in the form of 'space-time'. Both
of these papers are examples of incomplete conceptual understandings of the
nature of time as 'information'. In order to obtain any kind of consistency
and remain free of temporal paradoxes, 'time' must be understood in the
paradigm of the computational properties of matter in a dynamic universe.
The application of the relatively complex T-computer concept may not appeal
to adherents of the reductionist maxim of ''Ockham's Razor'', but if 'time'
were reducible to a simple equation or relationship within the frameworks of
'classical' or 20th century physics, there would be no 'problem of time' at
this time.

\section{Future Directions}

The ideas outlined in this paper are a sketch of an alternative way of
looking at the origins of 'time' implicit in the operational definitions and
devices (clocks) of everyday experience. T-computers at the quantum level
may someday be used to drive the information processing activities in
quantum computers, hybrid quantum-classical computers, and quantum
biological computers (e.g. 'photosynthetic' \cite{R15}). It may be that the
current use of system clocks driving large-scale classical computers is
unnecessary (see \cite{R16}). If local T-computers do their job of ordering
and time labeling information (where necessary) in such a way that non-local
entangled processing logic can 'instantaneously' perform calculations such
as those responsible for complex meta-infostates like consciousness, we may
see much faster classical computers as well as advances in quantum computer
architecture when the appropriate devices are finally engineered. The
designs for clockless chips in digital computers may foretell a new strategy
for computing in the quantum realm where information is propagated as simple
or complex infostates.

As for now, perhaps we will see that 'time' is something that living beings
construct to 'predict' how things change in our environment. We construct
'time' in order to understand evolutionary patterns and 'compute' their
trajectories into our 'future'. Time maps are essential in order to optimize
our survival strategies.

New and exciting possibilities will emerge from understanding the connection
between information and 'time' and the fundamental links to information
processing of various degrees of complexity in hierarchical systems.
Consciousness may have arisen as a survival tool based on our ability to
'time label' our world. Perhaps consciousness is the result of the evolution
of relatively simple T-computers in the single cell building blocks of the
hierarchical world of living creatures. Only 'time' (= consciousness?) will
tell.

\section{Acknowledgments}

I would like to thank V. A. Petrov and the Organizing Committee for inviting
me to share these ideas at the XXIV International Workshop on the
Fundamental Problems of High Energy Physics and Field Theory, Institute for
High Energy Physics, Protvino, Russia. I would also like to thank Arkadi
Lipkine and Shanti Goradia for valuable discussions and insights. I also
thank Toni for getting me to Moscow on 'time'!

\section{Captions for Figures}

Figure 1. A Feynman Diagram representation of proton-neutron scattering.

Figure 2. This illustrates a proton and neutron scattering interaction
creating a transient p-meson Feynman Clock (box) with a composite infostate,
'I', in 'info-space'.

Figure 3. This figure is a causal network node 'map' of the interaction of
'incoming' proton and neutron 'signals'. They 'collide' to create a
transient Feynman Clock (the 'force' mediating p-meson is 'inside' the FC).
The FC 'computed' (via conservation of momentum 'logic' of the transient FC)
outgoing signals follow new trajectories in space (vacuum). The trajectories
represent information flow in space. The calibration signals locate and
identify the particle signals and therefore the 'implied' FC reaction site
in space. These signals are sued by the T-computer to time label the
calibration signal event representing the spatial location of particles in
the transient causal network. The 'directions' of the calibration signals in
this figure are meant only to illustrate the ideas and have no special
orientation in space with respect to the particles under observation. Note
that the incoming calibration signals at the lower part of the figure are
not shown for simplicity. The time calibration signals at the detectors may
be part of a pulse train of signals from a cyclical standard clock. This
standard clock may also be coupled to the position detectors that process
the outgoing calibration signals of the incoming proton and neutron.

Figure 4. Schematic diagram of a simple idealized T-computer. The functions
of the logic 'gates' or nodes and the signals flowing between them represent
a general model of the computation of 'time differences' between events in a
'time-independent' information space of 'infostates', I(S), and the system
or causal network forming the time labeling computer is in state \^{e}S\~{n}%
. This state is the collective 'infostate' of the entire network forming the
T-computer including the active and inactive components of the relevant
computational network. The intent of modeling a T-computer is to show two
things. The first point is that 'time' is a computational artifact of a
signal mapping process that defines 'time' as function of the coincidence a
clock signal with an event signal. The second point is that the T-computer
is a component or sub-network of more complex information processing
systems. The T-computer is essentially a quantum computer with classical
'time' as an output. The principles in the signal mapping process also apply
for larger scale systems that can be treated by classical methods. The key
is that all the information used to define time at the macroscopic scale is
traceable backwards to the origin of information in the microscopic
'quantum' world.

Figure 5. This is a network node representation of the T-computer where the
flow of information between nodes and the infostates of the node are
indicated. The nodes are the logic gates of causal networks through which
information is transferred, modified and created as a result of the
interactions of signals with detectors and the signal processing logic of
physical systems such as atoms.

Figure 6. This illustrates the 'speeding up' and 'slowing down' of the
observers sense of external time based on the number of signals detected
from a standard clock in each of the three types of subjective 'time'
frames. The 'arrows of time' pointing from left to right, are the created by
the relationship between the standard clock, signals, and the observer
'components' acting together to form a T-computer in the observers reference
frame.


\section{Bibliography:}

\begin{thebibliography}{99}
\bibitem{R1}  \textbf{Time and Information: Part 1} by \emph{Scott M.
Hitchcock}, at the Los Alamos National Labs, LANL e-Print Archives report
number quant-ph/0012017, 12/6/00 and at NSCL as pre-print (MSUCL-1183).
www.nscl.msu.edu/news/nscl\_library/nscl\_preprint/MSUCL1183.pdf

\bibitem{R2}  \textbf{''The Direction of Time''} by \emph{Hans Reichenbach},
Edited by Maria Reichenbach, Dover Publications, Inc., Mineola, NY, 1999.

\bibitem{R3}  \textbf{Feynman Clocks, Casual Networks, and Hierarchical
Arrows of Time in Complex Systems From the Big Bang to the Brain} An invited
talk and paper given by \emph{Scott M. Hitchcock} at the 'XXIII
International Workshop on the Fundamental Problems of High Energy Physics
and Field Theory' at the Institute for High Energy Physics (IHEP), Protvino,
Russia, June 21-23, 2000. Published in the Proceedings of the Workshop and
at NSCL as pre-print (MSUCL-1172). It is also at the Los Alamos National
Labs, LANL e-Print Archives report number quant-ph/00100014.
www.nscl.msu.edu/news/nscl\_library/nscl\_preprint/MSUCL1172.pdf

\bibitem{R4}  \textbf{Quantum Clocks and the Origin of Time in Complex
Systems} by \emph{Scott M. Hitchcock}, Los Alamos National Lab, LANL e-Print
Archives report number gr-qc/9902046 v2, 20 February 1999, and NSCL Research
Report MSUCL-1123.
www.nscl.msu.edu/news/nscl\_library/nscl\_preprint/MSUCL1123.pdf

\bibitem{R5}  \textbf{Is There a 'Conservation of Information Law' for the
Universe?} by \emph{Scott M. Hitchcock} at the Los Alamos National Labs,
LANL e-Print Archives report number gr-qc/0108010, 3 Aug 2001,
http://xxx.lanl.gov/abs/gr-qc/0108010 and at NSCL as pre-print (MSUCL-1212).
Available at
www.nscl.msu.edu/news/nscl\_library/nscl\_preprint/MSUCL1212.pdf. \emph{%
There is some preliminary support that this proposed 'law' may have some
validity, at least for the special case of a Schwarzschild Black Hole. See
'Hawking radiation from decoherence'' by Claus Kiefer} at:
http://xxx.lanl.gov/abs/gr-qc/0110070

\bibitem{R6}  ''\textbf{Diagrammatica; The Path to ''Feynman Diagrams}'' by 
\emph{Martinus Veltman}, Cambridge University Press, 1995.

\bibitem{R7}  ''\textbf{A Guide to Feynman Diagrams in the Many-Body Problem}%
'' by \emph{Richard D. Mattuck}, Dover Publications, Inc., New York, 1992.

\bibitem{R8}  ''\textbf{Quantum Computation and Quantum Information''} by 
\emph{Michael A. Nielsen }and \emph{Isaac L. Chuang}, Cambridge University
Press, UK, 2000.

\bibitem{R9}  ''\textbf{Explorations in Quantum Computing'' }by \emph{Colin
P. Williams }and \emph{Scott H. Clearwater}, Springer-Verlag, New York, 1998.

\bibitem{R10}  ''\textbf{Lecture Notes for Physics 229: Quantum Information
and Computation''} by \emph{John Preskill}, California Institute of
Technology, September 1998, available at http://www.theory.caltech.edu/%
\symbol{126}preskill/ph229.

\bibitem{R11}  \textbf{The Early Universe as a Quantum Growing Network} by 
\emph{P. A. Zizzi}, the Los Alamos National Labs, LANL e-Print Archives
report number gr-qc/0103002, 6 Apr 2001,
http://xxx.lanl.gov/ftp/gr-qc/papers/0103/0103002.pdf

\bibitem{R12}  \textbf{Quantum computation in brain microtubules?} The
Penrose-Hameroff ''Orch OR'' model of consciousness. By \emph{Stuart Hameroff%
}, Philosophical Transactions Royal Society London (A) 356:1869-1896 (1998).

\bibitem{R13}  \textbf{''Computational capacity of the universe''} by \emph{%
Seth Lloyd}, at the Los Alamos National Labs, LANL e-Print Archives report
number quant-ph/0110141 v1, 24 Oct 2001,
http://xxx.lanl.gov/abs/quant-ph/0110141

\bibitem{R14}  \textbf{''Is there a real time ordering behind the nonlocal
correlations?'' }by \emph{Antoine Suarez}, at the Los Alamos National Labs,
LANL e-Print Archives report number quant-ph/0110124 v1, 20 Oct 2001,
http://xxx.lanl.gov/abs/quant-ph/0110124

\bibitem{R15}  \textbf{'Photosynthetic' Quantum Computers?} by \emph{Scott
M. Hitchcock} at the Los Alamos National Labs, LANL e-Print Archives report
number quant-ph/108087, 20 Aug 2001,
http://xxx.lanl.gov/abs/quant-ph/0108087 and at NSCL as pre-print
(MSUCL-1213).
www.nscl.msu.edu/news/nscl\_library/nscl\_preprint/MSUCL1213.pdf

\bibitem{R16}  \textbf{It's Time for Clockless Chips} by \emph{Claire
Tristram}, Technology Review, October 2001, Volume 104, Number 8, pp. 37-41.
http://www.technologyrevew.com/
\end{thebibliography}
\end{document}